\documentclass{mem}
\usepackage{natbib}
\usepackage{txfonts}
\usepackage{balance}
\usepackage{flushend}
\usepackage{graphicx}
\usepackage{comment}
\usepackage[breaklinks,pdftex]{hyperref}
\usepackage[separate-uncertainty=true,binary-units=true,per-mode=symbol]{siunitx}
\idline{75}{282}
\begin{document}

%%%%%%%%%%%%%%%%%%%%%%%%%%%%%%%%%%%%%%%%%%%%%%%%%%%
\newcommand{\astri}           {ASTRI}
\newcommand{\astrih}          {ASTRI-Horn}
\newcommand{\astriuno}        {ASTRI-1}
\newcommand{\miniarray}       {MiniArray}
\newcommand{\variance}        {Variance}
\newcommand{\var}             {VAR}
\newcommand{\scitech}         {\textit{SciTech}}
\newcommand{\vstar}           {\textit{V-Star}}
\newcommand{\runid}           {\cod{RUNID}}
\newcommand{\fov}             {FoV}
\newcommand{\psf}             {PSF}

\sisetup{
    range-phrase = \text{--},
    range-units = single,
}

% (tens of  \si{\arcsecond}) despite a modest camera resolution (few \si{\arcminute}).
% "ASTRI Project"
% "ASTRI-Horn prototype"
% "ASTRI-N telescope", when referring to the specific Nth telescope of the array (e.g., ASTRI-1)
% "ASTRI reduced array composed of k telescopes (ASTRI-kT)", where k = 2,3,4,5,6,7,8 (e.g., ASTRI-3T)
% "ASTRI array composed of 9 telescopes (ASTRI-9T)”, formerly known as ASTRI Mini-Array
%%%%%%%%%%%%%%%%%%%%%%%%%%%%%%%%%%%%%%%%%%%%%%%%%%%

\title{Learning from ASTRI-Horn:}
\subtitle{products and applications of Variance data}

\author{
S. Iovenitti\inst{1} 
\and S. Crestan\inst{2}
\and T. Mineo\inst{3}
\and G. Leto\inst{4}
\and A. Giuliani\inst{2}
\and S. Lombardi\inst{5}\\
%\bf{
for the ASTRI project*
}

\institute{ INAF - Osservatorio astronomico di Brera, via E. Bianchi 46, Merate (LC) 23807 - IT.
\and INAF - IASF, via Alfonso Corti 12, Milano (MI) 20133 - IT.
\and INAF - IASF, via Ugo La Malfa 153, Palermo (PA) 90146 - IT.
\and INAF - Osservatorio astrofisico di Catania, via Santa Sofia 78, Catania (CT) 95123 - IT.
\and INAF - Osservatorio astronomico di Roma, via Frascati 33, Mt. Porzio (RM) 00040 - IT.\\
(*) \href{http://www.astri.inaf.it/en/library/}{http://www.astri.inaf.it/en/library}\\
\email{simone.iovenitti@inaf.it}
}

\authorrunning{Iovenitti}
\titlerunning{Learning from ASTRI-Horn}

\date{Received: XX-XX-XXXX; Accepted: XX-XX-XXXX}

\abstract{
In the context of the \astri{} \miniarray{} project (9 dual-mirror air Cherenkov telescopes being installed at the Observatorio del Teide in the Canary Islands), the \astrih{} prototype was previously implemented in Italy (Sicily). It was a crucial test bench for establishing observation strategies, hardware upgrades, and software solutions. Specifically, during the winter 2022/2023 observing campaign, we implemented significant enhancements in using the so-called \variance{} mode, an auxiliary output of the ASTRI Cherenkov camera able to take images of the night sky background in the near-UV/visible band.  \variance{} data are now processed online and on site using a dedicated pipeline and stored in tech files. This data can infer possible telescope mis-pointing, background level, number of identified stars, and point spread function. In this contribution, we briefly present these quantities and their importance together with the algorithms adopted for their calculation. They provide valuable monitoring of telescope health and sky conditions during scientific data collection, enabling the selection of optimal time sequences for Cherenkov data reduction.
}

\maketitle{}

\section{Introduction}\label{sec:intro}
The ASTRI project (acronym of Astrofisica con Specchi a Tecnologia Replicante Italiana, \citealt{scuderi_astri_param}) is a significant Italian initiative aimed at advancing our understanding of very high-energy (VHE) gamma-ray astronomy (\citealt{ASTRI_SCIENCE_PILLARS}), up to \SI{300}{\tera\electronvolt} and beyond.
\begin{figure*}[t!]
\resizebox{\hsize}{!}{\includegraphics[clip=true]{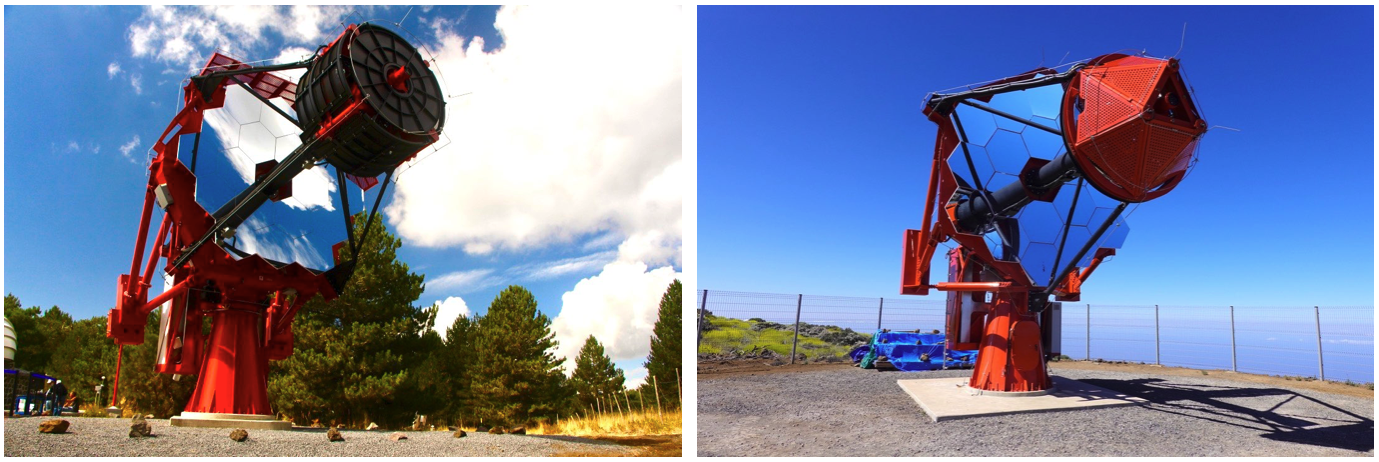}}
\caption{\footnotesize
The prototype telescope ASTRI-Horn on Mount Etna in Italy (\textit{left}) and ASTRI-1, the first Cherenkov telescope at the Teide Observatory in Canary Islands (\textit{right}).
}
\label{fig:astri}
\end{figure*}
During the past decade, ASTRI has pioneered the development of a new type of imaging atmospheric Cherenkov telescope (IACT, \citealt{IACTS_REVIEW}), characterized by a dual-mirror modified Schwarzschild-Couder configuration (\citealt{COUDER}).
This unique design, realized for the first time by ASTRI (\citealt{astri_optical_validation}), provides optical features that are essential to achieve the ambitious scientific objectives of the project. The optical system is aplanatic and isochronous, with a very wide field of view (\fov{}) of almost \SI{11}{\degree} and a quasi-flat response of the Point Spread Function (\psf{}, \citealt{ASTRI_validation_2019}). The scheme of the instrument presents a short focal length (f/0.5) and a primary mirror of \SI{4.3}{\meter} diameter, composed of 18 reflective segments, while the secondary mirror is monolithic. The telescope results in a very compact and robust mechanical structure (figure~\ref{fig:astri}).

The ASTRI Cherenkov camera (\citealt{ASTRI_camera_2018}) is installed at the focal point of the telescope. Its curved surface is equipped with miniaturized silicon photomultipliers (SiPM, \SI{7}{\milli\meter}), arranged into 37 tiles, each containing 64 sensors. The sky footprint of every pixel\footnote{
    The plate-scale is constant \citealt{astri_optical_validation}.
} is \SI{11}{\arcminute}, designed to match the size of the \psf{}.
The electronics of the camera is designed to image the atmospheric Cherenkov flashes on the timescale of few \si{\nano\second} (\citealt{ASTRI_CAMERA_ELECTRONICS}). Nevertheless, there is an ancillary output dedicated to the Night Sky Background (NSB), producing an image of the \fov{} every \SIrange{1}{3}{\second} in the frequency range of the instrument, i.e. between \SIrange{200}{500}{\nano\meter}. This is the so-called \variance{} (\var{}) data flow (\citealt{Segreto_calibration}), a gold mine of opportunities to calibrate and monitor the whole system using astrometry and photometry techniques (\citealt{TESI_SIMO_PHD}).

Currently, the ASTRI project is progressing towards the deployment of the ASTRI Mini-Array, a configuration of nine telescopes operating in stereoscopic mode at the Observatorio del Teide in the Canary Islands (\citealt{ASTRI_MA_TEIDE}).
Furthermore, the ASTRI structure will also be implemented for the Small-Sized Telescope (SST) of the Cherenkov Telescope Array Observatory (CTAO, \citealt{PARESCHI_SPIE2016}).
In this context, the prototype instrument \astrih{}, technology demonstrator and first telescope of the project, % (\citealt{ASTRI_CRAB_detection}),
has represented a vital testbed for refining calibration procedures, observational strategies, and developing software solutions. During the winter 2022/2023 observing campaign (\citealt{GAMMA24_CRESTAN}), significant improvements were made in the scientific exploitation of \variance{} data. In this contribution, we first provide a detailed examination of \var{} data products %quantities
and their significance, followed by a presentation of the current \var{} analysis use cases.

\section{\variance{} data products}
\variance{} data are images of the NSB produced in parallel to scientific data every $\sim$\SI{3}{\second}. For \astrih{} the limit magnitude is about $7^{\rm th}$ B mag and the \fov{} is reduced to \SI{7.8}{\degree} as only 21 PDM are installed at the
\begin{figure}[t]
\resizebox{\hsize}{!}{\includegraphics[clip=true]{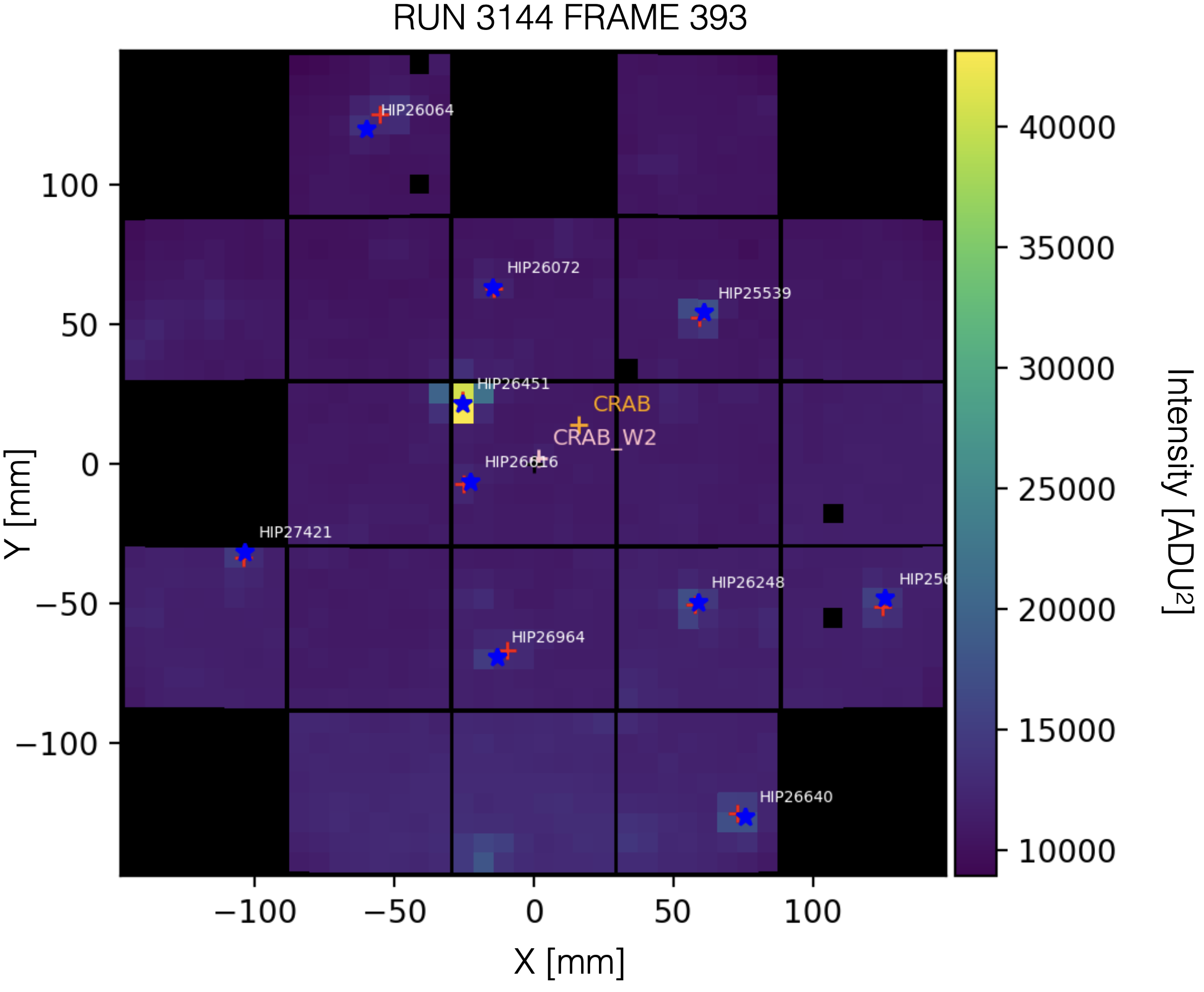}}
\caption{\footnotesize
\astrih{} \var{} image (21 PDM) processed with \vstar{} (\citealt{SPIE22_VSTAR}). Light spots (\textit{red}) are matched with stars (\textit{blue}) to retrieve the position of the source (\textit{yellow}) and the pointing coordinate in wobble mode (\textit{pink}).
}
\label{fig:var}
\end{figure}
focal plane (see figure~\ref{fig:var}, where unfortunately two PDMs are disconnected due to hardware problems in the last winter).
The importance of these images lies in the fact that they provide the only opportunity to inspect the \fov{} during regular data taking\footnote{While an optical camera is mounted at the focal plane during technical phases.}. In fact, \var{} images originate from the same optical system, detector, sky conditions, and time as the scientific data. Consequently, a fast variance analysis can provide crucial information not only for the calibration, monitoring, and data reduction of the system (section~\ref{sec:applications}), but also offer real-time feedback to the operators on shift during observations.
For this reason, we have set up a dedicated pipeline to process \var{} data both online and onsite. The output is a technical FITS file (the so-called \scitech{}) that aggregates several quantities, such as telescope mispointing, background level, the number of stars identified by the custom astrometry routine \vstar{}, and an estimation of the size of the focal spot. These metrics are essential for assessing the telescope's performance and the sky conditions during scientific observations, enabling the selection of optimal time sequences for Cherenkov data reduction by feeding the \scitech{} file into the Cherenkov data pipeline (\citealt{ASTRI_DPS}).

\begin{figure}[h]
\resizebox{\hsize}{!}{\includegraphics[clip=true]{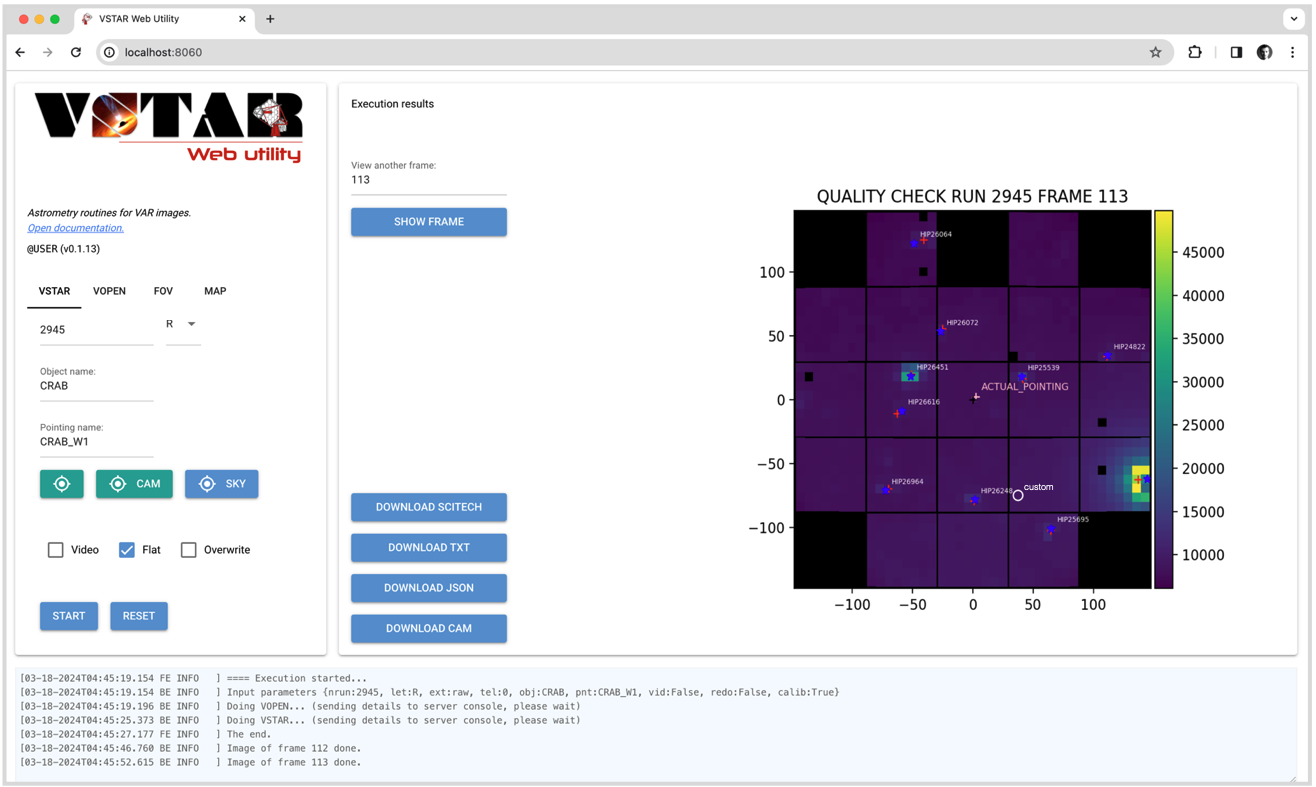}}
\caption{\footnotesize
Web user interface for the astrometry utilities developed for the ASTRI telescope.
The new version of \vstar{} spots also planets in the \fov{}, as in the case of Mars in the \var{} image on the right.
}
\label{fig:gui}
\end{figure}

\begin{figure}[t]
\centering
\resizebox{0.8\hsize}{!}{\includegraphics[clip=true]{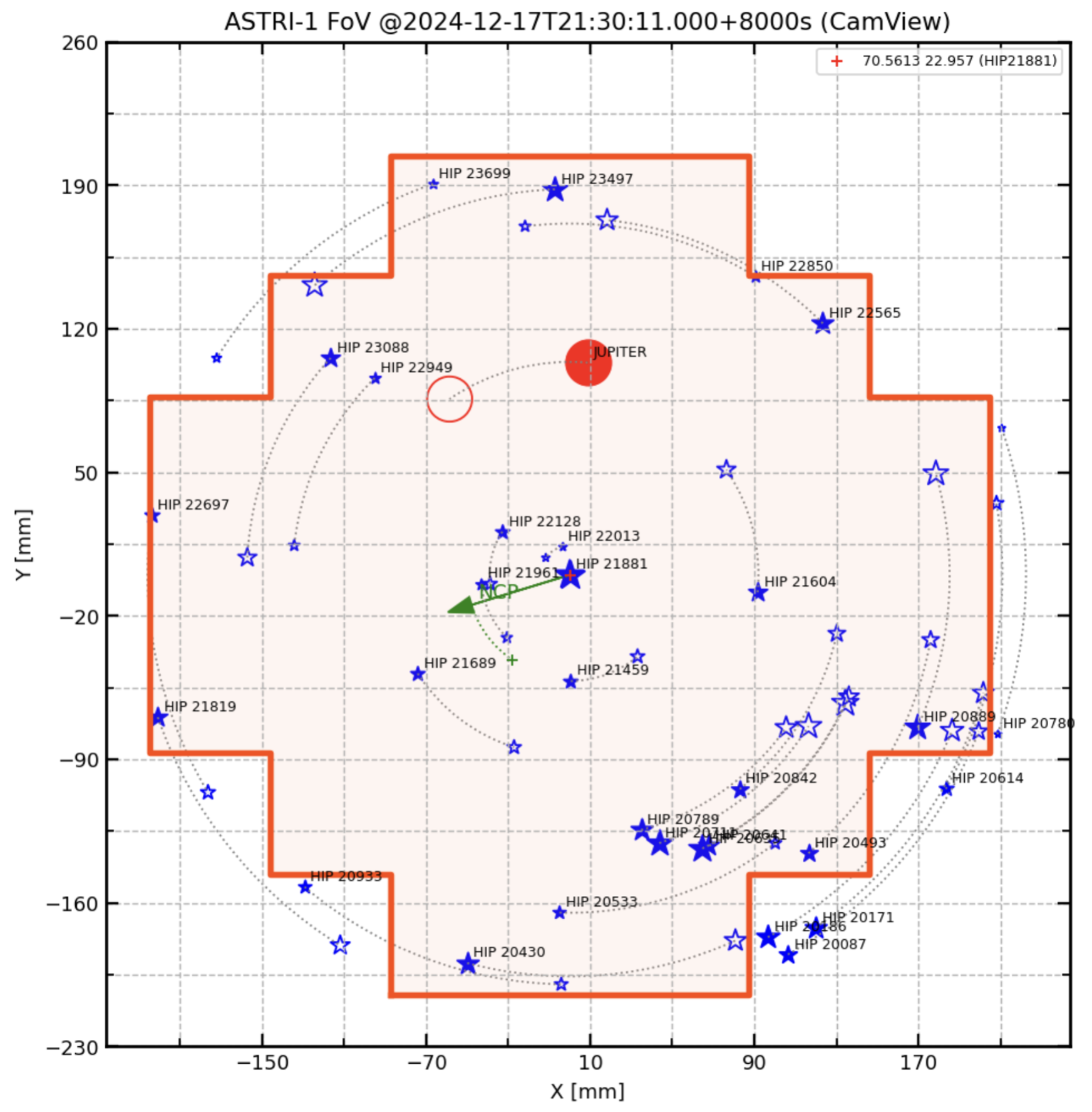}}
\caption{\footnotesize
Example of the \fov{} simulator. An acquisition of \SI{8000}{\second} with ASTRI-1, close to Jupyter.}
\label{fig:fov}
\end{figure}

To provide fast and user-friendly access to \var{} data, we developed the graphical user interface (GUI) shown in figure~\ref{fig:gui}. It is written in Python using the NiceGui library\footnote{\href{https://nicegui.io/}{https://nicegui.io/}}, and it is accessible through an internet browser with a VPN connection.
Several utilities are integrated into this GUI, including \vstar{}, a \fov{} simulator (figure~\ref{fig:fov}, \citealt{IOVE_ICRC21_STARCOVERAGE}), and a sky map generated with Aladin Lite (\citealt{ALADIN_V3}). The GUI supports both quick-look data review and online analysis, and enables the visualization of any custom Right Ascension/Declination object over the \var{} image (white circle), as well as the inverse coordinates transformation (camera-to-sky).

\section{Applications of VAR analysis}\label{sec:applications}
\var{} data are crucial for both scientific analysis and technical operations. Here, we present the current status of the most significant use cases of the \variance{}, focusing on the calibration and monitoring of the system.

\subsection{Calibration}\label{sec:calibration}
Every ASTRI telescope is equipped with a Pointing Monitoring Camera (PMC), an auxiliary device mounted on the back of the secondary mirror and dedicated to the measurement of the telescope sky position with high precision ($\sim$\SI{0.25}{\arcsecond}).
Despite this, the Cherenkov camera may not be perfectly aligned with the PMC, due to gravity flexures or mechanical tolerances.
Consequently, to ensure the pointing accuracy of scientific data it is crucial to characterize the possible offset between the two devices.
\begin{figure}[b]
\resizebox{\hsize}{!}{\includegraphics[clip=true]{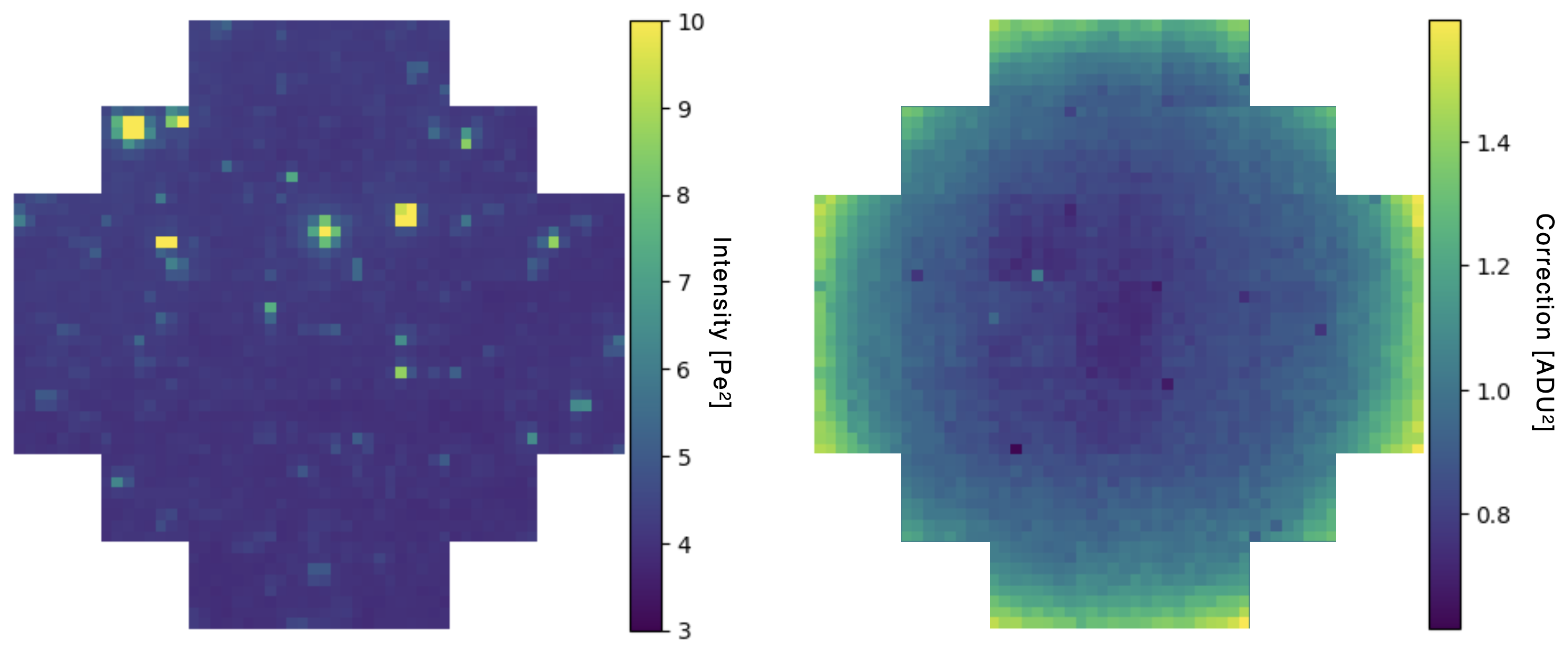}}
\caption{\footnotesize
A \var{} image of \astriuno{} (\textit{left}) corrected with the  the flat-field
dithering-like matrix (\textit{right}).}
\label{fig:flat}
\end{figure}
With \astrih{} we developed an effective strategy for this calibration, exploiting the \fov{} rotation recorded by both the instruments (\citealt{IOVENITTI_CAMERA_ALIGNMENT}).
At the \miniarray{}, this offset is included in the pointing model and not inserted as a correction at the level of data reduction.
Moreover, the relative rotation of the cameras is also characterized, using the transit of bright stars at the local meridian. %(\citealt{IOVE_ICRC21_VARIANZA})

\begin{figure}[h]
\resizebox{\hsize}{!}{\includegraphics[clip=true]{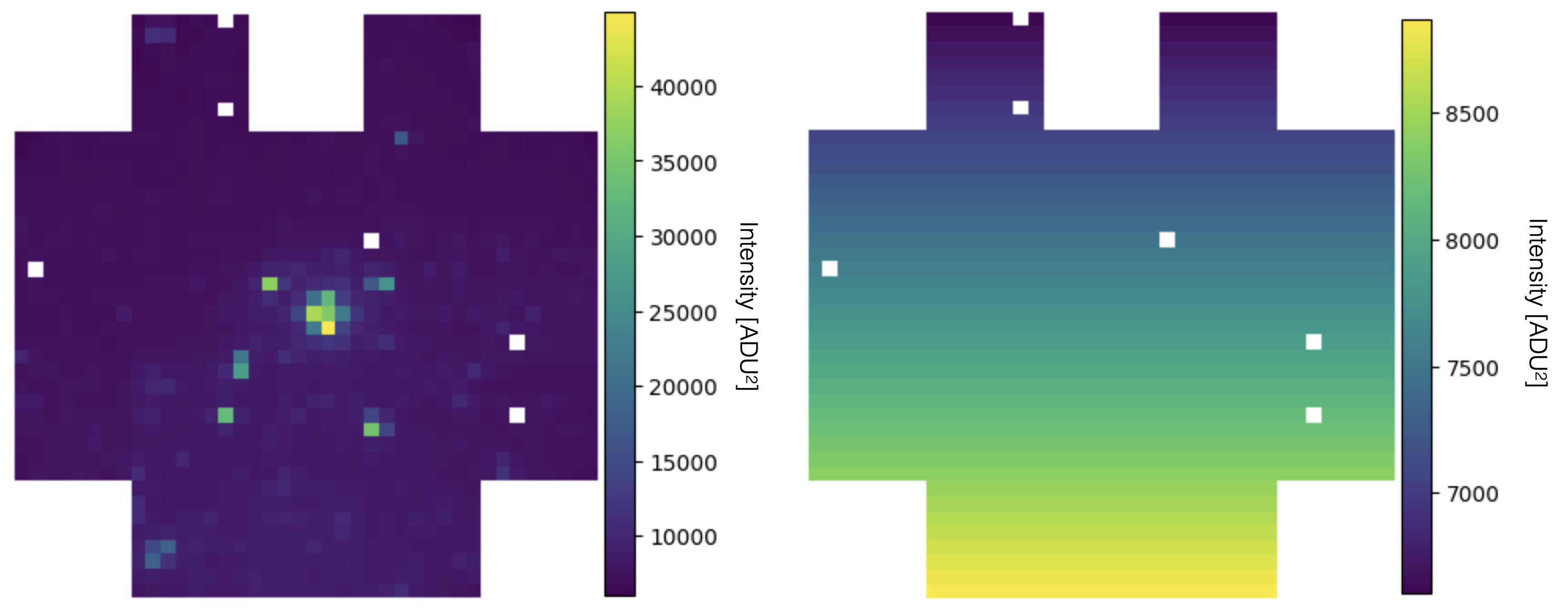}}
\caption{\footnotesize
A real \astrih{} \var{} image (Vega region, \textit{left})
and its linear flat-field gradient (\textit{right}).
Six pixels and a PDM were not connected.
}
\label{fig:gradient}
\end{figure}

The \variance{} is also used for the flat-field calibration of the \fov{}.
First, the equalization of the pixel response is achieved using the NSB with the camera detached from the telescope (see \citealt{CONTINO_FLAT_FIELD}). Second, with the camera at the focal point, the entire ASTRI imaging system can be flat-fielded using a dithering-like procedure (figure~\ref{fig:flat}).%, \citealt{DITHERING_STRATEGY}).
Lastly, the gradient due to the horizon is removed, fitting each single image (figure~\ref{fig:gradient}).

\subsection{Monitoring}\label{sec:monitoring}
Due to the coarse angular resolution of the ASTRI camera (\SI{11}{\arcminute}), individual \var{} images do not provide a detailed representation of the night sky.
However, they enable the monitoring of several quantities that are otherwise difficult to access. For this reason, one of the first steps in the data reduction process is the astrometric calibration of variance images, performed using \vstar{}.
This software allows us to reconstruct the sky coordinates of the Cherenkov camera center with modest precision ($\sim$\SI{2}{\arcminute}), but high rate ($\sim$\SI{1}{\hertz}), making it suitable for real-time quick monitoring (\citealt{SPIE22_OOQS}).
%\footnote{ Implemented in the Online Observation Quality System (OOQS, \citealt{SPIE22_OOQS}). }.
During the last observing campaign with \astrih{}, this software was 
redesigned and enhanced. It now identifies planets in the \fov{} (figure~\ref{fig:gui}), tracks the position of selected objects and issues warnings for bright lights in the \fov{}\footnote{
    e.g. airplanes, satellites, car flashes, etc.
}.It also operates effectively for fixed alt-az pointings and under high-background conditions, such as during bright moon phases or in regions with significant light pollution.%~\citealt{NSB_CATANIA}
The NSB level is recorded in the \scitech{} files, and serves as a metric for evaluating the quality of an observing run, as stray light can contaminate Cherenkov shower images or cause spurious triggers
Currently, the NSB is also monitored by dedicated ancillary instruments (as the SQM or UVSiPM described in \citealt{GAMMA24_MINEO}), but the \variance{} is the only one to use the ASTRI optical system. For this reason, over long timescales, variance data can be used to monitor the degradation of the mirrors' integrated reflectivity\footnote{
    Which is a critical aspect in our system.
}.
Additionally, \var{} data can be used to monitor the size of the focal spot in time. Although it is impossible to measure accurately the optical \psf{} due to the large camera pixels, \variance{} can detect if it changes\footnote{
    Ice or strong wind could move mirror segments.
}
considering the average number of pixels illuminated by every star.
This ensures the correct imaging of Cherenkov showers.

% Optimization of optical parameters in Monte Carlo productions
% filter cutoff (red stars intensity)

\section{Conclusions} \label{sec:conclusion}
In addition to conducting scientific observations, the prototype telescope \astrih{} has served as a critical facility for development and testing. This contribution outlines a suite of tools and procedures specifically designed for the calibration and monitoring of the system. These tools leverage the \variance{} method, which has proven essential for enhancing the scientific accuracy of the telescope. Building on the lessons learned with \astrih{}, these tools have been optimized and fine-tuned, making them ready for deployment in the upcoming ASTRI \miniarray{}.

%offer valuable insights into the operational health of the telescope and the quality of the observational data, thereby enhancing the overall efficacy of the ASTRI project.

\begin{acknowledgements}
This work was conducted in the context of the ASTRI Project. We gratefully acknowledge support from the people, agencies, and organisations listed here: \href{http://www.astri.inaf.it/en/library/}{http://www.astri.inaf.it/en/library}. This paper went through the internal ASTRI review process. 
\end{acknowledgements}

\bibliographystyle{aa}
\bibliography{bibliography}

\begin{thebibliography}{21}
\expandafter\ifx\csname natexlab\endcsname\relax\def\natexlab#1{#1}\fi

\bibitem[{Baumann {et~al.}(2022)Baumann, Boch, Pineau, Fernique, Bot, \&
  Allen}]{ALADIN_V3}
Baumann, M., Boch, T., Pineau, F.-X., {et~al.} 2022, Astronomical Society of
  the Pacific Conference Series, 532, 7

\bibitem[{Catalano {et~al.}(2018)Catalano, Capalbi, Gargano, Giarrusso,
  Impiombato, Rosa, Maccarone, Mineo, Russo, Sangiorgi, Segreto, Sottile,
  Biondo, Bonanno, Garozzo, Grillo, Marano, Romeo, Scuderi, Canestrari,
  Conconi, Giro, Pareschi, Sironi, Conforti, Gianotti, \&
  Gimenes}]{ASTRI_camera_2018}
Catalano, O., Capalbi, M., Gargano, C., {et~al.} 2018, in Ground-Based and
  {{Airborne Instrumentation}} for {{Astronomy VII}}, Vol. 10702 ({SPIE}),
  1070237

\bibitem[{Contino {et~al.}(2023)Contino, Catalano, Mollica, Capalbi, Corpora,
  Gargano, Iovenitti, Sangiorgi, Sottile, Leto, Zanmar~Sanchez, Mineo,
  Pareschi, Scuderi, \& {ASTRI project}}]{CONTINO_FLAT_FIELD}
Contino, G., Catalano, O., Mollica, D., {et~al.} 2023, in Proceedings of 38th
  ICRS — {PoS} (Nagoya, Japan: Sissa Medialab), 739

\bibitem[{Couder(1926)}]{COUDER}
Couder, A. 1926, Comptes Rendus, 183, 1276

\bibitem[{Crestan {et~al.}(2024)}]{GAMMA24_CRESTAN}
Crestan, S. {et~al.} 2024, in this proceedings

\bibitem[{{de Naurois} \& Mazin(2015)}]{IACTS_REVIEW}
{de Naurois}, M. \& Mazin, D. 2015, Comptes Rendus Physique, 16, 610

\bibitem[{Giro {et~al.}(2019)Giro, Canestrari, Bruno, Catalano, Fugazza,
  Palombara, Maccarone, Pareschi, Russo, Scuderi, Segreto, Sironi, Tosti,
  Marchiori, Busatta, Marcuzzi, \& Folla}]{ASTRI_validation_2019}
Giro, E., Canestrari, R., Bruno, P., {et~al.} 2019, in Optics for {{EUV}},
  {{X-Ray}}, and {{Gamma-Ray Astronomy IX}}, Vol. 11119 ({InSPIE}), 111191E

\bibitem[{Giro {et~al.}(2017)Giro, Canestrari, Sironi, Antolini, Conconi,
  Fermino, Gargano, Rodeghiero, Russo, Scuderi, Tosti, Vassiliev, \&
  Pareschi}]{astri_optical_validation}
Giro, E., Canestrari, R., Sironi, G., {et~al.} 2017, Astronomy \& Astrophysics,
  608, A86

\bibitem[{Iovenitti(2021)}]{IOVE_ICRC21_STARCOVERAGE}
Iovenitti, S. 2021, in Proceedings of 37th {{ICRC}} --- {{PoS}}({{ICRC2021}}),
  Vol. 395 (SISSA Medialab), 735

\bibitem[{Iovenitti(2022)}]{TESI_SIMO_PHD}
Iovenitti, S. 2022, PhD thesis, Universit{\`a} degli Studi di Milano,
  Dipartimento di Fisica

\bibitem[{Iovenitti {et~al.}(2021)Iovenitti, Sironi, Giro, Segreto, Catalano,
  \& Capalbi}]{IOVENITTI_CAMERA_ALIGNMENT}
Iovenitti, S., Sironi, G., Giro, E., {et~al.} 2021, Experimental Astronomy, 53,
  117

\bibitem[{Iovenitti {et~al.}(2022)Iovenitti, Sironi, Mineo, Capalbi, Catalano,
  Lombardi, Scuderi, Segreto, \& Sottile}]{SPIE22_VSTAR}
Iovenitti, S., Sironi, G., Mineo, T., {et~al.} 2022, in {{SPIE Conference
  Proceedings}}

\bibitem[{Lombardi {et~al.}(2022)Lombardi, Lucarelli, Bigongiari, Gallozzi,
  Cardillo, Mastropietro, Saturni, Visconti, Antonelli, Bulgarelli, Capalbi,
  Catalano, Compagnino, Conforti, Crestan, Cusumano, D'aì, Germani, Giuliani,
  Iovenitti, Parola, Maccarone, Mineo, Mollica, Pagliaro, Parmiggiani, Perri,
  Pintore, Scuderi, Tosti, Vercellone, \& Zampieri}]{ASTRI_DPS}
Lombardi, S., Lucarelli, F., Bigongiari, C., {et~al.} 2022, in Software and
  {Cyberinfrastructure} for {Astronomy} {VII}, Vol. 12189 (SPIE), 275--291

\bibitem[{Mineo {et~al.}(2024)}]{GAMMA24_MINEO}
Mineo, T. {et~al.} 2024, in this proceedings

\bibitem[{Pareschi(2016)}]{PARESCHI_SPIE2016}
Pareschi, G. 2016, in Ground-Based and {{Airborne Telescopes VI}}, Vol. 9906
  (SPIE), 1992--2004

\bibitem[{Parmiggiani {et~al.}(2022)Parmiggiani, Bulgarelli, Baroncelli, Addis,
  Fioretti, Catalano, Conforti, Fiori, Gianotti, Iovenitti, Lucarelli,
  Maccarone, Mineo, Lombardi, Pastore, Russo, Sangiorgi, Tosti, Trifoglio, \&
  Zampieri}]{SPIE22_OOQS}
Parmiggiani, N., Bulgarelli, A., Baroncelli, L., {et~al.} 2022, Proc. SPIE, 12

\bibitem[{Scuderi(2018)}]{scuderi_astri_param}
Scuderi, S. 2018, in Ground-Based and {{Airborne Telescopes VII}}, Vol. 10700
  ({InSPIE}), 107005Z

\bibitem[{Scuderi {et~al.}(2022)Scuderi, Giuliani, Pareschi, Tosti, Catalano,
  Amato, Antonelli, Becerra~Gonz{\`a}les, Bellassai, Bigongiari, Biondo,
  Boettcher, Bonanno, Bruno, Bulgarelli, Canestrari, Capalbi, Cardillo,
  Conforti, Contino, Corpora, Costa, Cusumano, D'A{\`i}, {de Gouveia Dal Pino},
  Della~Ceca, Escribano~Rodriguez, {Falceta-Gon{\c c}alves}, Fermino, Fiori,
  Fioretti, Fiorini, Gallozzi, Gargano, Garozzo, Germani, Ghedina, Gianotti,
  Giarrusso, Gimenes, Giordano, Grillo, Grivel~Gelly, Impiombato, Incardona,
  Incorvaia, Iovenitti, La~Barbera, La~Palombara, La~Parola, Lamastra, Lessio,
  Leto, Lo~Gerfo, Lodi, Lombardi, Longo, Lucarelli, Maccarone, Marano,
  Martinetti, Mereghetti, Miccich{\'e}, Millul, Mineo, Morlino, Morselli,
  Naletto, Nicotra, Pagliaro, Parmiggiani, Piano, Pintore, Poretti, Olmi,
  Rodeghiero, Rodriguez~Fernandez, Romano, Romeo, Russo, Sangiorgi, Saturni,
  Schwarz, Sciacca, Sironi, Sottile, Stamerra, Tagliaferri, Testa, Umana,
  Uslenghi, Vercellone, Zampieri, \& Zanmar~Sanchez}]{ASTRI_MA_TEIDE}
Scuderi, S., Giuliani, A., Pareschi, G., {et~al.} 2022, Journal of High Energy
  Astrophysics

\bibitem[{Segreto {et~al.}(2019)Segreto, Catalano, Maccarone, Mineo,
  La~Barbera, \& Cta Astri~Project}]{Segreto_calibration}
Segreto, A., Catalano, O., Maccarone, M.~C., {et~al.} 2019, in Proceeding of
  {{Science}}, Vol.~36 (SISSA Medialab), 791

\bibitem[{Sottile {et~al.}(2016)}]{ASTRI_CAMERA_ELECTRONICS}
Sottile, G. {et~al.} 2016, Proc. SPIE, 9906, 99063D

\bibitem[{Vercellone(2021)}]{ASTRI_SCIENCE_PILLARS}
Vercellone, S. 2021, PoS, ICRC2021, 896

\end{thebibliography}

\end{document}